\documentclass[preprintnumbers,showpacs,amsmath,amssymb,floatfix,9pt,prd,onecolumn,
superscriptaddress,nofootinbib]{revtex4}
\usepackage{CJK}
\usepackage{graphicx}
\usepackage{latexsym}
\usepackage{epsfig}
\usepackage{enumerate}
\usepackage{amssymb}
\usepackage[utf8]{inputenc}
\begin{document}
\title{ Noncommutative wormhole in de Rham-Gabadadze-Tolley like massive gravity}

\author{Piyali Bhar}
\email{piyalibhar90@gmail.com}
\affiliation{Department of
Mathematics,Government General Degree College Singur, Hooghly, West Bengal 712 409,
India}

\author{Allah Ditta}
\email{mradshahid01@gmail.com}
\affiliation{Department ofMathematics, Shanghai University and Newtouch Center for Mathematics of Shanghai University, Shanghai,200444,PR China}

\author{Abdelghani Errehymy}
\email{abdelghani.errehymy@gmail.com}
\affiliation{Astrophysics Research Centre, School of Mathematics, Statistics and Computer Science, University of KwaZulu-Natal, Private Bag X54001,
Durban 4000, South Africa}
\date{\today}
\begin{abstract}
The wormhole solution in dRGT massive gravity is examined in this paper in the background of non-commutative geometry. In order to derive the wormhole model, along with the zero tidal force, we assume that the matter distribution is given by the Gaussian and Lorentzian distributions. The shape function in both models involves the massive gravity parameters $m^2c_1$ and $m^2c_2$. But the spacetime loses its asymptotic flatness due to the action of the massive gravity parameter. It is noticed that the asymptotic flatness is affected by the repulsive effect induced in the massive gravitons that push the spacetime geometry very strongly. We observed that each model violates the null energy criteria, indicating the presence of exotic matter which is necessary to sustain the wormholes. The exotic matter is measured using the volume integral quantifier. Moreover, it is discovered that the model is stable under the hydrostatic equilibrium condition by utilizing the TOV equation. Finally, our research encompassed an exploration of the repulsive influence exerted by gravity. Our findings demonstrated that the presence of repulsive gravity results in a negative deflection angle for photons following null geodesics. Remarkably, we consistently observed negative values for the deflection angle across all values of $r_0$ in the two scenarios examined. This consistent negativity unequivocally signifies the manifestation of the repulsive gravity effect.
\end{abstract}


\maketitle

\section{Introduction}
The general theory of relativity (GR) formulated by Einstein has been remarkably successful in elucidating various phenomena in astrophysics and cosmology. However, the discovery of the late-time acceleration of the universe, attributed to dark energy, and the existence of dark matter have presented challenges to the prevailing understanding based on GR \cite{SupernovaSearchTeam:1998fmf,SupernovaCosmologyProject:1998vns,Hanany:2000qf,deBernardis:2001mjh,Peebles:2002gy,SDSS:2003tbn,Riess:2006fw,Amanullah:2010vv,WMAP:2010qai}. These findings have raised important questions and cast doubts on the continued success of Einstein's theory of gravitation. It is worth noting that one of the notable solutions to Einstein's field equations, known as the Schwarzschild solution, was derived by K. Schwarzschild in 1916 \cite{Schwarzschild:1916uq}. This solution describes a static and spherically symmetric black hole. The confirmation of gravitational waves (GWs) \cite{LIGOScientific:2016aoc} has provided compelling evidence for the existence of stellar-mass black holes in the natural universe. Interestingly, as far back as 1916, Ludwing Flamm \cite{Flamm:1916aoc} recognized that Einstein's equations have another solution, now known as a white hole. Unlike black holes, white holes were initially believed to expel matter and light from their event horizon. These two solutions could represent distinct regions in spacetime that are connected by a conduit, which later became known as a ``bridge''. In 1935, Albert Einstein and Nathan Rosen used the principles of general relativity to propose the existence of such ``bridges'' in spacetime \cite{Einstein:1935tc}. Several decades later, in 1957, the term \textit{wormhole} was coined by Misner and Wheeler \cite{Misner:1957mt}, leading to a renewed focus on studying these intriguing phenomena. However, it was later discovered that the original version of wormholes is not traversable due to the rapid opening and closing of their throat. This means that they are not practical for passage or communication.

To avoid the rapid closure of the wormhole's throat, a potential solution is to introduce a scalar field that interacts with gravity. This additional component gives rise to a broader category of wormholes, which were initially proposed by Ellis \cite{Ellis:1973yv} and independently by Bronnikov \cite{Bronnikov:1973fh}. However, one of the major obstacles in the study of wormholes is the need for exotic matter to maintain their stability and structure. Exotic matter is characterized by violating the energy conditions that are commonly observed in conventional physics. In 1988, Morris and Thorne introduced certain conditions for wormholes to be traversable, which can be found in \cite{Morris:1988cz}. These solutions are derived by considering a peculiar type of matter, known as exotic matter, which possesses unusual properties necessary for maintaining the structure of the wormhole. Notably, this exotic matter exhibits negative energy density, satisfying the flare-out condition while violating the weak energy condition \cite{Morris:1988cz,Morris:1988tu}.

It has been established that modified theories of gravity, such as $f(R)$ theory, have a noteworthy contribution in the study of wormholes. Lobo and Oliveira conducted a study \cite{Lobo:2009ip} on traversable wormholes within the framework of $f(R)$ gravity. They examined the factors responsible for violating the null energy condition and supporting the existence of wormholes. They also investigated wormhole solutions for various shape functions. Another work by Bronnikov et al. \cite{Bronnikov:2010tt} discussed the presence of wormholes in scalar-tensor theory and $f(R)$ gravity. The null and weak energy conditions for wormholes with constant shape and redshift functions in $f(R)$ gravity can be found in \cite{Saeidi:2011zz}. S. Bahamonde et al. \cite{Bahamonde:2016ixz} developed dynamical wormholes in $f(R)$ gravity, while Kuhfittig \cite{Kuhfittig:2018vdg} explored wormhole solutions using different types of shape functions in the same framework. Lemos et al. \cite{Lemos:2003jb} explored the study of static and spherically symmetric traversable wormholes in the presence of a cosmological constant, while Errehymy et al. \cite{Errehymy:2023rnd,Errehymy:2024yey} addressed the topic of wormholes surrounded with and without dark matter halos.

A diverse array of wormhole solutions can be found within modified theories of gravity. For instance, wormholes have been studied in Einstein-Gauss-Bonnet gravity \cite{Maeda:2008nz}, $f(R, \Phi)$ gravity \cite{Zubair:2017oir}, as well as theories involving both $f(R)$ and $f(R, T)$ \cite{DeFalco:2023twb,Errehymy:2024spg}. Additionally, wormhole solutions have been explored in Born-Infeld gravity \cite{Shaikh:2018yku}, Eddington-inspired Born-Infeld gravity \cite{Harko:2013aya}, and even in the context of noncommutative geometry \cite{Jamil:2013tva,Rahaman:2012pg}. Mustafa et al. \cite{Mustafa:2022fxn} used observational data in the matter coupling gravity formalism in $f(R,\,T)$ gravity to determine the potential of generalized wormhole development due to dark matter in the galactic halo. Using the Karmarkar condition in the f(Q) gravity formalism-where Q is the nonmetricity scalar-authors in ref.~ \cite{Mustafa:2021ykn} achieved wormhole solutions. They demonstrate how the combination of these components offers the prospect of obtaining traversable wormholes that satisfy the energy conditions. New wormhole solutions in the fourth-order modified Ricci inverse gravity background are the subject of the research work by Mustafa \cite{Mustafa:2023mls}. By displaying the valid region for the majority of the wormhole geometry under the influence of the relevant parameters, two new classes of wormhole solutions are analyzed in this paper. In the extended f(R,Lm) gravity, Mustafa et al. \cite{Mustafa:2024iwu} used an anisotropic matter source and a particular type of energy density demonstrating cold dark matter halo and quantum wave dark matter halo to calculate two different wormhole solutions. The properties of the exotic matter within the wormhole geometry and the matter contents via energy conditions are studied in detail in this paper. Another interesting work on wormholes can be found in \cite{Mustafa:2021vqz}. Moreover, there are numerous other models of wormholes beyond those mentioned above, as seen in works such as \cite{Boehmer:2012uyw,KordZangeneh:2015dks,Mehdizadeh:2017tcf,Jusufi:2018waj,Ovgun:2018xys,Golchin:2019qch,Dai:2019mse,Ditta:2021uoe,Mustafa:2024jsv,Errehymy:2024cgy,Errehymy:2023rsm,Waseem:2023ejh,Mustafa:2023ojf,Mustafa:2024ark}. These encompass the construction of traversable wormholes within various types of modified gravity theories. Einstein \cite{Einstein:1936llh} was the first to explore the field of GR, and one of its early and valuable discoveries was gravitational lensing. This phenomenon occurs when an object that is very massive bends incoming light in the same way as a lens, providing viewers with more information about the source of the light. Following the observational confirmation of light's deflection, as suggested theoretically, interest in the field of gravitational lensing grew \cite{Dyson:1920cwa,Eddington:1920cwa}. This phenomenon provides opportunities to investigate exoplanets, dark matter, and dark energy, reaching beyond the solar system. It was first identified observationally as a double quasar system in 1979 \cite{Walsh:1979nx}. Since then, it has become one of the important research topics in gravitational physics. The first successful astrometric microlensing measurement of a white dwarf's mass, Stein 2051 B, marked a significant milestone \cite{Sahu:2017ksz}. One fascinating feature of gravitational lensing is that, in situations with intense lensing, it may bend light eternally in unstable light rings, producing a multitude of relativistic pictures \cite{Virbhadra:1999nm,Bozza:2001xd,Bozza:2002zj}. Strong and weak versions of this phenomena are useful analytical tools for studying gravitational fields around different cosmic objects, such as wormholes and black holes. Gravitational lensing has been widely used in theoretical and astrophysical studies to investigate wormholes, demonstrating the importance of this phenomenon in current research \cite{Tsukamoto:2016qro,Shaikh:2017zfl,Jusufi:2017mav,Ovgun:2018fnk,Javed:2019qyg,Ovgun:2020yuv,Javed:2022fsn}.

In this study, we investigate a noncommutative wormhole in the context of de Rham-Gabadadze-Tolley (dRGT)-like massive gravity. The investigation into massive gravity commenced before the discovery of the accelerated expansion of the cosmos. In 1939, the authors \cite{Fierz:1939r} introduced a linear theory of massive gravity, which treated the graviton as a mass term. However, this proposed model had some drawbacks \cite{Zakharov:1970cc,vanDam:1970vg}, as the asymptotic-massless limits of the linear theory did not align with the predictions of GR. In 1972, the author \cite{Vainshtein:1972sx} put forward the idea that the nonlinear approach in the theory of massive gravity could potentially address the problem. However, this approach gave rise to a new issue known as the Boulware-Deser ghost \cite{Hinterbichler:2011tt}. Subsequently, the Boulware-Deser ghost was successfully eliminated through the introduction of a new nonlinear version of massive gravity proposed by de Rham, Gabadadze, and Tolley (dRGT) \cite{deRham:2010kj}. Since then, the dRGT formulation has become a notable framework for studying the universe on a cosmic scale. Furthermore, there have been numerous intriguing articles exploring the applications of dRGT massive gravity to exotic objects such as black holes \cite{Ghosh:2015cva,Boonserm:2017qcq}.

The paper followed a structured organization, beginning with Section \ref{s2} offering a concise overview of massive gravity. This was followed by Section \ref{s3}, which delved into the exploration of wormhole solutions. In Section \ref{s4}, the study focused on analyzing the photon deflection angle on null geodesics and investigating the deviation of photons' paths. Finally, Section \ref{s5} encompassed discussions and conclusions drawn from the findings presented in the paper.

\section{Brief review of massive gravity}\label{s2}
Here in this section, we describe the field equations for dRGT model of massive gravity, by using the spherically symmetric space time. As stated distinctly in the references \cite{1,I21}, we here start with our study with the massive gravity dynamics equivalent to the dRGT model, using the matter Lagrangian $S_{\text {matter }}$:
\begin{eqnarray}\label{1}
    S_G&=&\frac{1}{16 \pi G} \int d^4 x \sqrt{-g}\left[R+m^2 \sum_i^4 c_i U_i(g, f)\right]+S_{\text {matter }},
\end{eqnarray}
in this instance, $R$ is the Ricci scalar curvature, $m$ is a graviton mass-related variable, and $G$ is the Newton constant of gravitation. The constants $c_i$, the metric tensor $g$, and a fixed symmetric tensor $f$ are examples of integral components, and show the e dynamical and reference metrics,
respectively. The effective potential of a graviton, $U_i$, can be determined in four dimensions of spacetime as follows:
\begin{eqnarray}\label{2}
\begin{cases}
 U_1&=[\mathcal{K}], \\
 U_2&=[\mathcal{K}]^2-\left[\mathcal{K}^2\right], \\
 U_3&=[\mathcal{K}]^3-3[\mathcal{K}]\left[\mathcal{K}^2\right]+2\left[\mathcal{K}^3\right], \\
 U_4&=[\mathcal{K}]^4-6\left[\mathcal{K}^2\right][\mathcal{K}]^2+8\left[\mathcal{K}^3\right][\mathcal{K}]+3\left[\mathcal{K}^2\right]^2-6\left[\mathcal{K}^4\right],
 \end{cases}
\end{eqnarray}
in this instance, $[\mathcal{K}]$ represents the trace of the metric $\mathcal{K}_v^\mu$, and $\mathcal{K}v^\mu=\sqrt{g^{\mu \lambda} f{\lambda v}}$. Changing the action (\ref{1}) with regard to $g_v^\mu$ yields the equation of motion controlling this gravitational system:
\begin{eqnarray}\label{3}
    R_{\mu \nu}-\frac{1}{2} R g_{\mu \nu}+m^2 \chi_{\mu \nu}=\frac{8 \pi G}{c^4} T_{\mu \nu},
\end{eqnarray}
where
\begin{eqnarray}\label{4}
\begin{cases}
\chi_{\mu \nu}&=-\frac{c_1}{2}\left(U_1 g_{\mu \nu}-\mathcal{K}_{\mu \nu}\right)-\frac{c_2}{2}[U_2 g_{\mu \nu}-2 U_1 \mathcal{K}_{\mu \nu}\\&+2 \mathcal{K}_{\mu \nu}^2]-\frac{c_3}{2}\big[U_3 g_{\mu \nu}-3 U_1 \mathcal{K}_{\mu \nu}+6 U_1 \mathcal{K}_{\mu \nu}^2\\&-6 \mathcal{K}_{\mu \nu}^3\big]-\frac{c_4}{2}\big[U_4 g_{\mu \nu}-4 U_3 \mathcal{K}_{\mu \nu}+12 U_2 \mathcal{K}_{\mu \nu}^2\\&-24 U_1 \mathcal{K}_{\mu \nu}^3+24 \mathcal{K}_{\mu \nu}^4\big].
\end{cases}
\end{eqnarray}
In this case, we present a spherically symmetric metric as:
\begin{eqnarray}\label{5}
    d s^2=-e^{2 \phi(r)} d t^2+e^{2 \lambda(r)} d r^2+r^2\left(d \theta^2+\sin \theta^2 d \phi^2\right),
\end{eqnarray}
here, radial functions are denoted by $\lambda(r)$ and $\phi(r)$. The particular form of the reference metric that we have chosen is as follows:
\begin{eqnarray}\label{6}
    f_{\mu \nu}=\operatorname{diag}\left(0,0, C^2, C^2 \sin ^2 \theta\right),
\end{eqnarray}
where $C$ denotes the positive constant. The distinct functional forms of $U_i$ can be obtained using the metric ansatz (\ref{6}) and are as follows:
\begin{eqnarray}\label{7}
\begin{cases}
     U_1&=\frac{2 C}{r}, \quad U_2=\frac{2 C^2}{r^2}, \\
 U_i&=0, \quad i>2 .
 \end{cases}
\end{eqnarray}

Moreover, the energy-momentum tensor used for an anisotropic fluid is \cite{3}:
\begin{eqnarray}\label{8}
    T_{\mu \nu}=\left(\rho+p_t\right) u_\mu u_v+p_t g_{\mu \nu}+\left(p_r-p_t\right) n_\mu n_v,
\end{eqnarray}
where $\rho$ denotes energy density and $p_{\mathrm{r}}$ and $p_t$ denote radial and transverse pressure, respectively. The four-velocity is denoted by $u_{\mu}$, and the four-radial vector by $n_\mu$. The field equations' non-zero components can be acquired by using the energy-momentum tensor from Eq. (\ref{8}) and the metric given in Eq. (\ref{5}) as follows:

\begin{eqnarray}
\frac{8 \pi G}{c^2} r^2 \rho&=&e^{-2 \lambda}\left(2 r \lambda^{\prime}-1\right)+1+m^2 C\left(c_2 C+c_1 r\right),\label{9} \\
\frac{8 \pi G}{c^4} r^2 p_r&=&e^{-2 \lambda}\left(2 r \phi^{\prime}+1\right)-1-m^2 C\left(c_2 C+c_1 r\right), \label{10}\\
\frac{8 \pi G}{c^4} r p_t&=&e^{-2 \lambda}\left(\phi^{\prime}+r \phi^{\prime 2}-\lambda^{\prime}-r \lambda^{\prime} \phi^{\prime}+r \phi^{\prime \prime}\right)-\frac{m^2 c_1 C}{2},\label{11}
\end{eqnarray}
 where $'$ denotes the derivative with respect to radial coordinate $r$. In the next section, we are going to solve the eqns. (\ref{9})-(\ref{11}) under certain conditions to obtain the model of the wormhole in this modified gravity.
\section{Wormhole Solutions}\label{s3}
In eqn.~(\ref{5}), $\phi(r)$ is referred to as the redshift function because of its relationship to the gravitational redshift and is taken to be finite everywhere to prevent event horizons which makes the wormhole (WH) traversable. $e^{-2\lambda}=1-\frac{b(r)}{r}$, $b(r)$, on the other hand, represents the shape function of the wormhole. The radial coordinate r extends to $\infty$ from a minimum value $r_0$, where $r_0$ represents the throat of the WH and it satisfies the formula $b(r_0)=r_0$.\\
For the sake of physical admissibility, the wormhole shape function $b(r)$ needs to adhere to the following conditions:
\begin{itemize}
    \item \textbf{Throat condition:} At the throat $r_0$, $b(r_0)=r_0$ and for $r>r_0$, $1-\frac{b(r)}{r}>0$.
    \item \textbf{Flaring-out condition:} At the throat, the radial differential of the shape function $b'(r)$ should be less than $1$.
    \item  \textbf{Asymptotic Flatness condition:} If $r\rightarrow\infty$, then $\frac{b(r)}{r}\rightarrow0$.
    \item   \textbf{Proper radial distance function:}  In traversable wormhole the proper radial distance function $l(r)=\pm \int (\frac{r-b(r)}{r})^{-\frac{1}{2}}dr$ must be finite everywhere in the domain. Thus, $b(r)$ should be less than $r$ for all $r$. The sign $\pm$ represents the upper and lower universes.
\end{itemize}
\subsection{Model I with Gaussian distribution}
In non-commutative geometry, smeared objects take the place of point-like particles. Space-time is encoded in the commutator $[x^{\mu},x^{\nu}]=i\theta^{\mu\nu}$, $\theta^{\mu\nu}$ being the anti-symmetric matrix. It causes the fundamental cell discretization of spacetime, like how the phase space is discretized by the Planck constant $\hbar$. We shall take into consideration the Gaussian distribution as a noncommutative energy density profile to describe our current model in this subsection, which has the following expression \cite{Nicolini:2005vd}:
\begin{eqnarray}\label{rho1}
\rho = \frac{M}{(4\pi\theta)^{\frac{3}{2}}}e^{-\frac{r^2}{4\theta}},
\end{eqnarray}
where the source's total mass is $M$ and $\theta$ is the non commutative parameter. Mass $M$ of the source is not perfectly localized at the point, but rather it is distributed throughout a region of linear size $\sqrt{\theta}$. The coordinate commutator's inherent uncertainty is responsible for this. This results from the inherent uncertainty of the coordinate commutator.\\
Along with this, to solve the field equations, let us assume a constant redshift function, i.e., $\phi$=constant.\\
Assuming $G=c=1$ and using the expression of $\rho$ given in (\ref{rho1}), we solve the equation (\ref{9}) for $e^{-2\lambda}$ which is obtained as,
\begin{eqnarray}
e^{-2\lambda}&=&1 + C m^2 \left(C c_2 + \frac{c_1 r}{2}\right)+\frac{2M}{\sqrt{\theta \pi}}e^{-\frac{r^2}{4\theta}}-\frac{2M}{r}\text{erf}\left(\frac{r}{2\sqrt{\theta}}\right)+\frac{E_1}{r},
\end{eqnarray}
where $E_1$ is the constant of integration. `erf(z)' is the ``error function" that arises in integrating the normal distribution. It has the following expression:
\begin{eqnarray*}
    erf(z)=\frac{2}{\sqrt{\pi}}\int_0^z e^{-t^2}dt,
\end{eqnarray*}
and in Maclaurin series expansion, it can be written as,
\begin{eqnarray*}
     erf(z)&=&\frac{2}{\sqrt{\pi}}\left[z-\frac{z^3}{3}+\frac{z^5}{10}-\frac{z^7}{42}+...\right].
\end{eqnarray*} \\
Now using the relationship $e^{-2\lambda}=1-\frac{b(r)}{r}$, the shape function of the wormhole is obtained as,
\begin{eqnarray}\label{sh1}
b(r)=-\frac{1}{2} C m^2 r (2 C c_2 + c_1 r) - \frac{2Mr}{\sqrt{\pi \theta}}e^{-\frac{r^2}{4\theta}}+2M\text{erf}\left(\frac{r}{2\sqrt{\theta}}\right)-E_1.
\end{eqnarray}
The value of the integration constant can be obtained from the formula $b(r_0)=r_0$.
The radial and transverse pressure are obtained as,
\begin{eqnarray}
p_r&=&\frac{M}{4r^2 \pi^{\frac{3}{2}}\sqrt{\theta}}e^{-\frac{r^2}{4\theta}}+\frac{ 2 E_1 - c_1 C m^2 r^2- 4 M \text{erf}\left(\frac{r}{2\sqrt{\theta}}\right)}{16\pi r^3},\\
p_t&=&\frac{M}{8\pi r^3}\text{erf}\left(\frac{r}{2\sqrt{\theta}}\right)-\frac{1}{32\pi r} C m^2c_1-\frac{M(r^2 + 2 \theta)}{16r^2(\pi \theta)^{\frac{3}{2}}}e^{-\frac{r^2}{4\theta}}-\frac{E_1}{16\pi r^3}.
\end{eqnarray}
\begin{figure}[htbp]
    \centering
    \includegraphics[scale=.75]{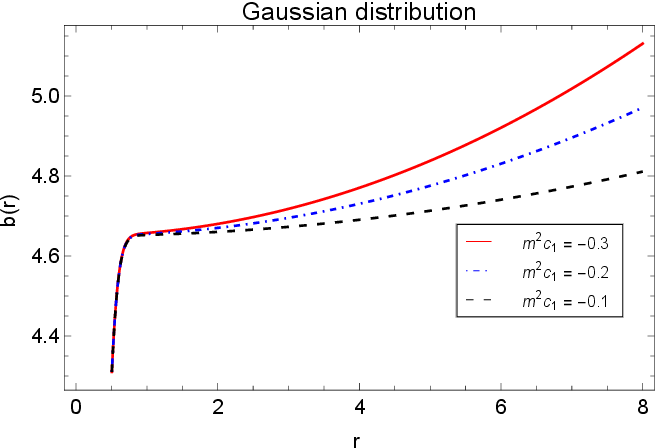}
    \includegraphics[scale=.75]{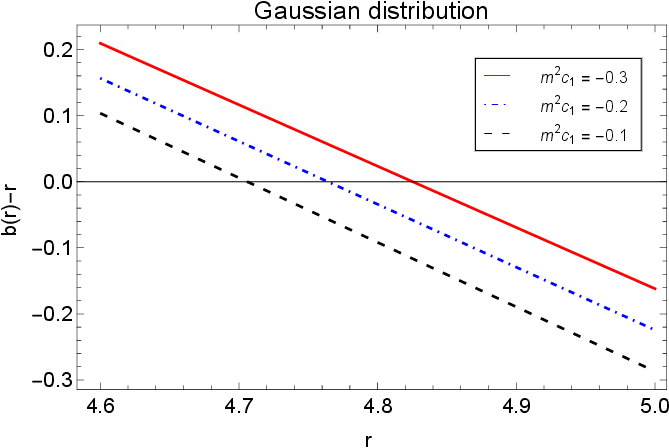}\\
    \includegraphics[scale=.75]{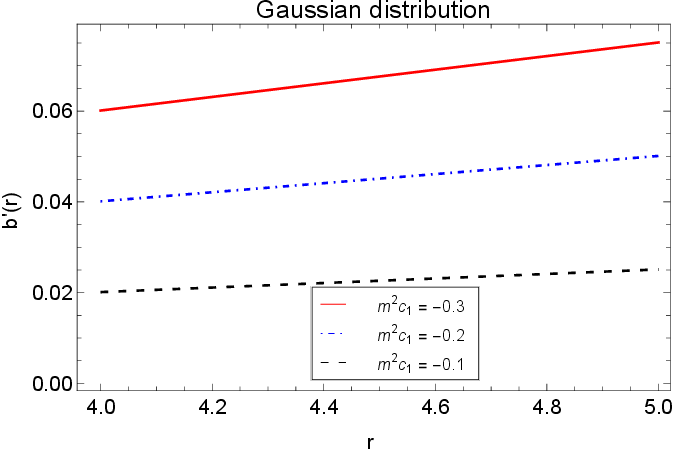}
    \includegraphics[scale=.85]{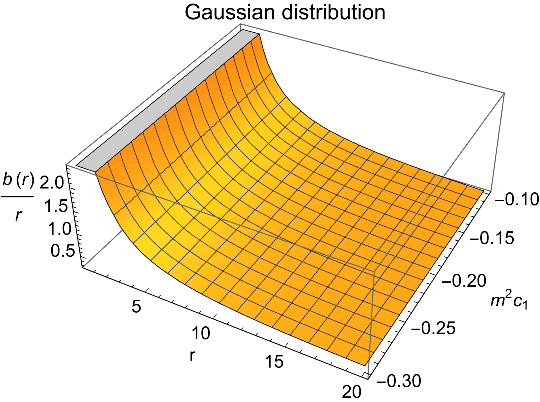}
        \caption{Different properties of the shape function with Gaussian distribution. For drawing the plots the values of the parameters are taken as follows: $C = 0.05,\, m^2c_2= -0.05,\,M = 1.8,\,\theta = 0.02$ and $E_1 = -1.05$.}
        \label{fig:1}
\end{figure}

\subsection{Model II with Lorentzian distribution}
We draw inspiration for the second wormhole model from the work of Mehdipour \cite{Mehdipour:2011mc}. Consequently, we suppose that the particle-like gravitational source has a Lorentzian distribution, and as a result, the energy density profile is as follows:
\begin{eqnarray}\label{rho2}
    \rho = \frac{\mathcal{M} \sqrt{\psi}}{\pi^2 (r^2 + \psi)^2},
\end{eqnarray}
where $\mathcal{M}$ and $\psi$ respectively denote the smeared mass distribution and the noncommutativity parameter.\\
Now by assuming the constant redshift function, we solve the field equations (\ref{9})-(\ref{11}) with the help of the density function given in eqn. (\ref{rho2}) which gives,
\begin{eqnarray}
    e^{-2\lambda}&=&1+\mathcal{M}^2C(c_2C+\frac{c_1}{2}r)-\frac{8\mathcal{M}\sqrt{\psi}}{r\pi}\left[-\frac{r}{2 (r^2 + \psi)} +\frac{1}{2\sqrt{\psi}} tan^{-1}\left(\frac{r}{\sqrt{\psi}}\right)\right]+\frac{E_2}{r},
\end{eqnarray}
here, $E_2$ represents the constant of integration. Using the well-known relationship between the shape function and the metric coefficient $e^{-2\lambda}$, the shape function of the wormhole in this model is obtained as,
\begin{eqnarray}\label{sh2}
b(r)&=&\frac{8\mathcal{M}\sqrt{\psi}}{\pi}\left[-\frac{r}{2 (r^2 + \psi)} +\frac{1}{2\sqrt{\psi}} tan^{-1}\left(\frac{r}{\sqrt{\psi}}\right)\right]-\frac{1}{2} C \mathcal{M}^2 r (2 C c_2 + c_1 r)-E_2,
\end{eqnarray}
The integration constant $E_2$ can be obtained from the relationship $b(r_0)=r_0$.\\
We consequently obtain the radial and transverse pressure as follows:
\begin{eqnarray}
    p_r&=&\frac{8 \mathcal{M} r \sqrt{\psi} + 2 E_2 \pi (r^2 + \psi) -
 c_1 C \mathcal{M}^2 \pi r^2 (r^2 + \psi) -
 8 \mathcal{M} (r^2 + \psi) tan^{-1}\left(\frac{r}{\sqrt{\psi}}\right)}{16 \pi^2 r^3 (r^2 + \psi)},\\
 p_t&=&\frac{8 \mathcal{M} (r^2 + \psi)^2 tan^{-1}\left(\frac{r}{\sqrt{\psi}}\right)-2 E_2 \pi (r^2 + \psi)^2 -
 r (c_1 C \mathcal{M}^2 \pi r (r^2 + \psi)^2 - 8 \mathcal{M} \sqrt{\psi} (3 r^2 + \psi))}{32 \pi^2 r^3 (r^2 + \psi)^2}
\end{eqnarray}

\begin{figure}[htbp]
    \centering
    \includegraphics[scale=.75]{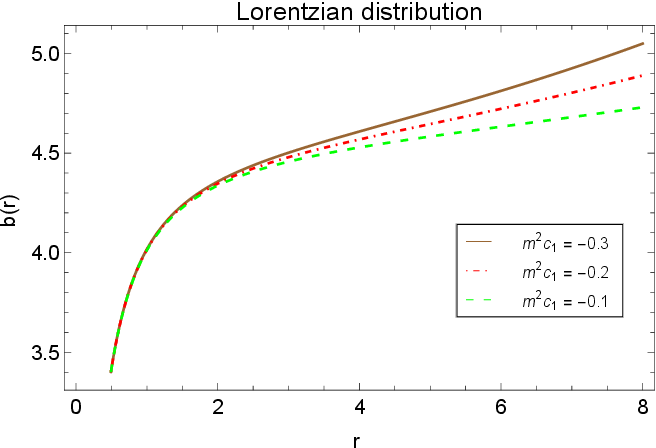}
    \includegraphics[scale=.75]{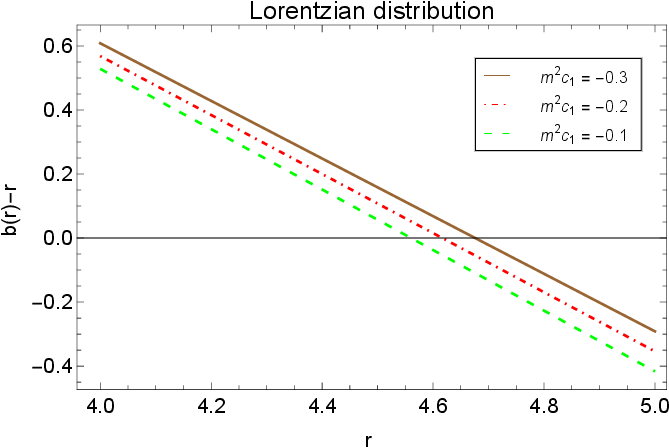}\\
    \includegraphics[scale=.75]{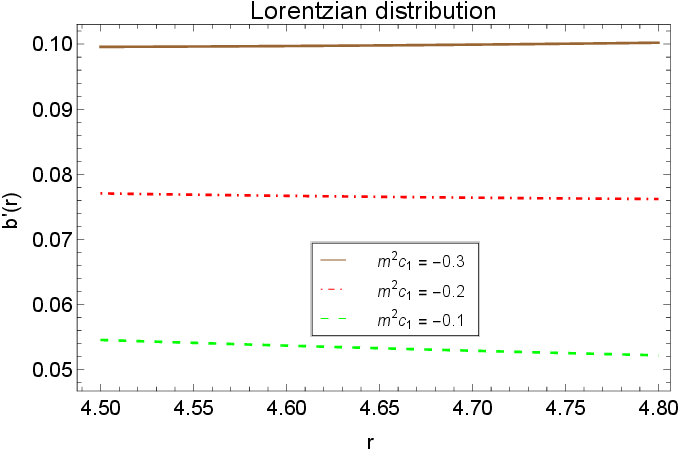}
    \includegraphics[scale=.8]{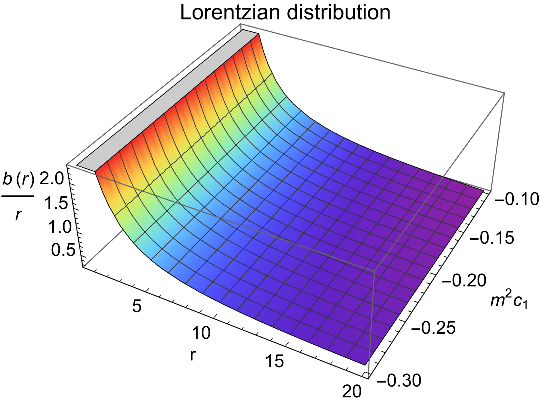}
        \caption{Different properties of the shape function with Lorentzian distribution. For drawing the plots we have taken $C= 0.05,\, m^2c_2 =-0.05,\,\mathcal{M} = 1.8,\, \psi = 0.02$ and $E_2 = -1.05$}
        \label{fig:2}
\end{figure}
We are going to discuss the shape function that we have obtained and the conditions that must be met for a wormhole to exist. To accomplish this, we carefully choose the appropriate free parameters which are mentioned in Figs.~\ref{fig:1}-\ref{fig:2}. Assuming a constant redshift function, we examine how the shape functions related to the Gaussian distribution as well as the Lorentzian distribution behave. The shape function and the flaring out condition for three different values of $m^2c_1$ are shown in Figs.~\ref{fig:1}-\ref{fig:2}. The shape function clearly shows positive increasing behavior. Nevertheless, the shape function shows a declining trend as the value of the model parameter $m^2c_1$ increases for both the two cases. Moreover, the upper right panels of Figs.~\ref{fig:1}-\ref{fig:2} show the behavior of $b(r)-r$, where the graphs cut the `r-axis' the position of the throat located at that point. By assuming the suitable values of the model parameters, the position of the throats for both the wormholes are listed in Tables~\ref{tb9}-\ref{tb10} for three different values of $m^2c_1$.\\
Furthermore, as can be seen from the bottom panels of Figs.~\ref{fig:1}-\ref{fig:2}, at the throat of the wormhole, the flaring out requirement, $b^{'}(r)<1$ for three different values of $m^2c_1$, is met. Additionally, the asymptotic behavior of the shape function over a range of $m^2c_1$ is shown in the right-hand side of the bottom panels of Fig.~\ref{fig:1}-\ref{fig:2}. It shows that the ratio $\frac{b(r)}{r}$ does not approach zero as $r~\rightarrow~\infty$, and therefore the asymptotic flatness condition is violated for both cases.

\begin{table*}[t]
\centering
\caption{The position of the throat of the wormhole for Gaussian distribution with $C = 0.05,\, m^2c_2= -0.05,\,M = 1.8,\,\theta = 0.02$ and $E_1 = -1.05$.}\label{tb9}
\begin{tabular}{@{}ccccccccccccc@{}}
\hline
$m^2c_1$&& $r_0$(in km.)  \\
\hline
-0.1&& 4.706\\
-0.2&& 4.764\\
-0.3&& 4.825\\
\hline
\end{tabular}
\end{table*}

\begin{table*}[t]
\centering
\caption{The position of the throat of the wormhole for Lorentzian distribution with $C= 0.05,\, m^2c_2 =-0.05,\,\mathcal{M} = 1.8,\, \psi = 0.02$ and $E_2 = -1.05$}\label{tb10}
\begin{tabular}{@{}ccccccccccccc@{}}
\hline
$m^2c_1$&& $r_0$(in km.)  \\
\hline
-0.1&& 4.56\\
-0.2&& 4.617\\
-0.3&& 4.676\\
\hline
\end{tabular}
\end{table*}

\subsection{Visualization of wormhole}
We next look over the embedding surface diagram and determine which conditions are necessary in order to represent the embedded wormhole configuration. Using $\theta=2 \pi$ and $t=$ const. in Eq. (\ref{5}), we can take specified spherical symmetric space-time with an equatorial slice, which in turn is as follows:
\begin{eqnarray}\label{ad1}
    d s^2=\left(1-\frac{b(r)}{r}\right)^{-1} d r^2+r^2 d \phi^2,
\end{eqnarray}
One can embed Eq. (\ref{ad1}) into 3-dimensional cylindrically symmetric Euclidean spacetime as follows:
\begin{eqnarray}\label{ad2}
    d s_{\Xi}^2=d z(r)^2+d r^2+r^2 d \phi^2 .
\end{eqnarray}
The above Eq. (\ref{ad2}) can be rewritten as
\begin{eqnarray}\label{ad3}
    d s_{\Xi}^2=\left(1+\left(\frac{d z(r)}{d r}\right)^2\right) d r^2+r^2 d \phi^2.
\end{eqnarray}
\begin{figure}[htbp]
    \centering
     \includegraphics[scale=.7]{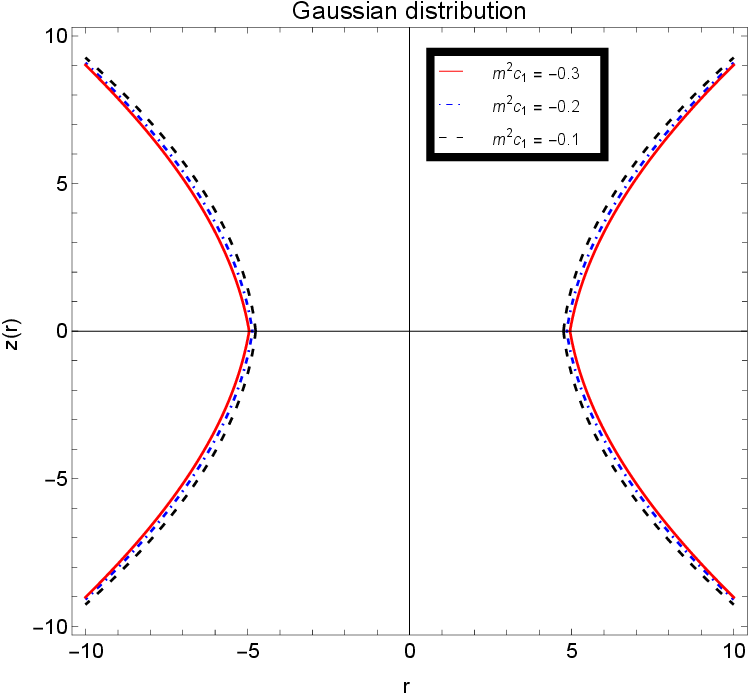}
    \includegraphics[scale=.7]{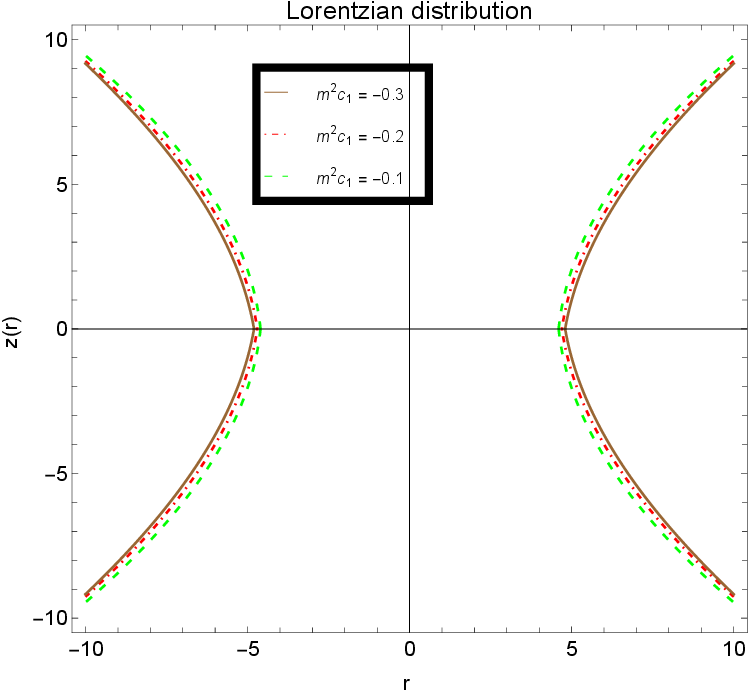}
        \caption{The embedding diagrams of the wormholes are shown. The values of the parameters are the same as Fig.~\ref{fig:1}-\ref{fig:2}. }
        \label{z1}
\end{figure}
\begin{figure}[htbp]
    \centering
     \includegraphics[scale=.4]{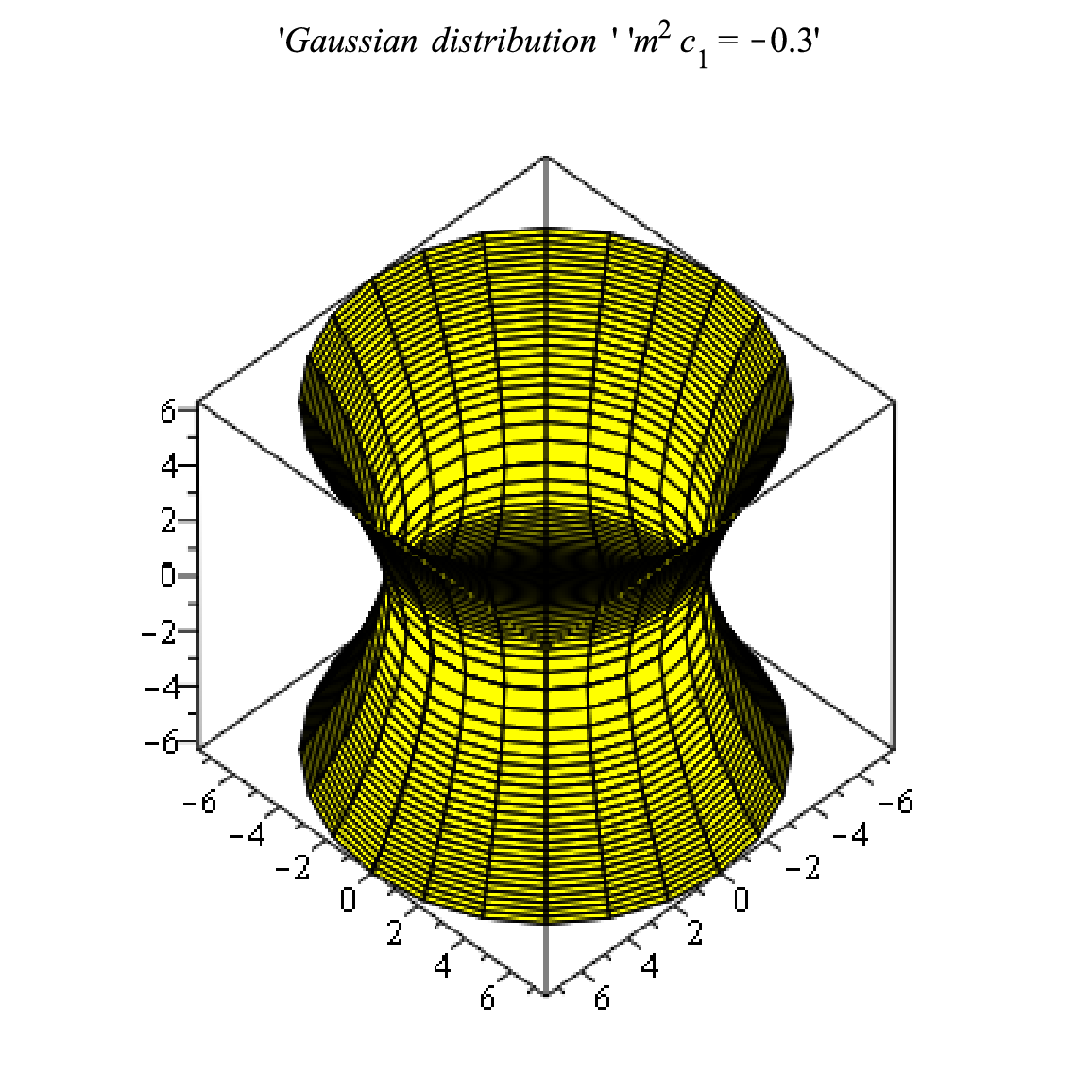}
    \includegraphics[scale=.4]{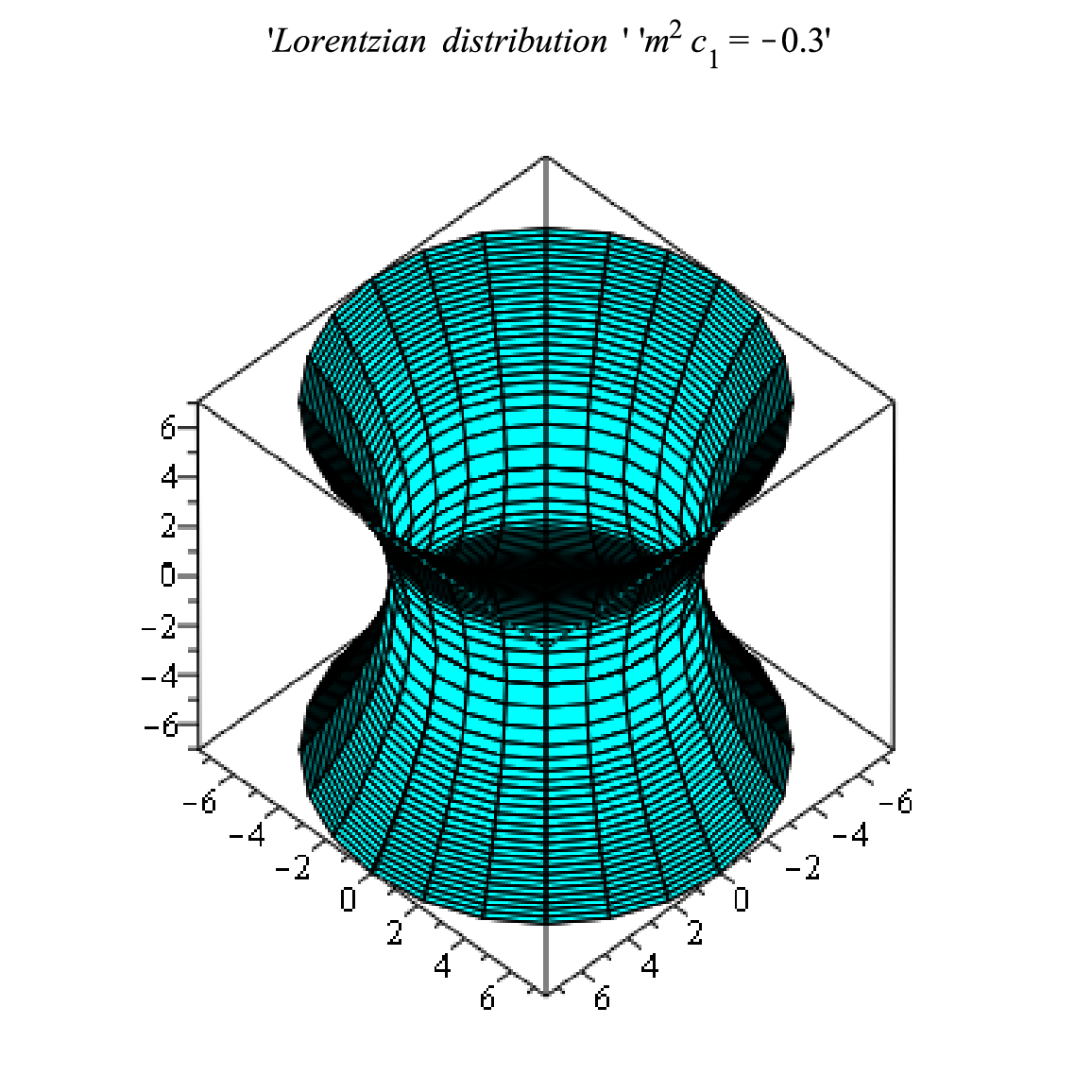}
        \caption{The full visualization of the wormholes is shown. The values of the parameters are the same as Fig.~\ref{fig:1}-\ref{fig:2}.}
        \label{z2}
\end{figure}
The relationship that results from matching Eqs. (\ref{ad1}-\ref{ad3}) is as follows:
\begin{eqnarray}
    \frac{d z(r)}{d r}= \pm\left(\frac{r}{b(r)}-1\right)^{-1 / 2}.
\end{eqnarray}
Fig.~\ref{z1} displays the embedded surface diagram of the wormhole for three different values of $m^2c_1$. In Fig.~\ref{z2}, the full visualization of the wormhole is obtained by revolving the Fig.~\ref{z1} with respect to `z-axis'.

\subsection{Energy conditions}
\begin{figure}[htbp]
    \centering
     \includegraphics[scale=.6]{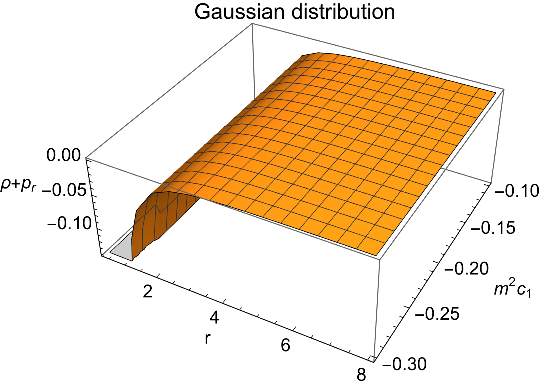}
    \includegraphics[scale=.6]{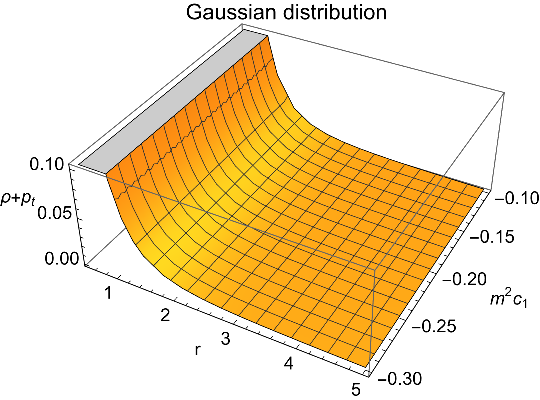}
    \includegraphics[scale=.6]{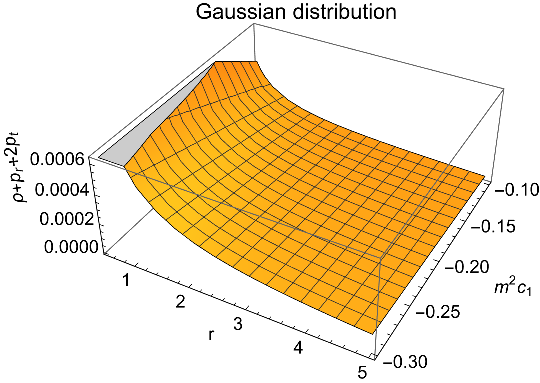}
    \includegraphics[scale=.6]{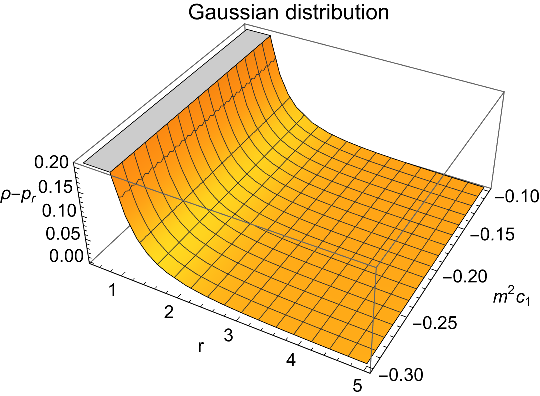}
    \includegraphics[scale=.6]{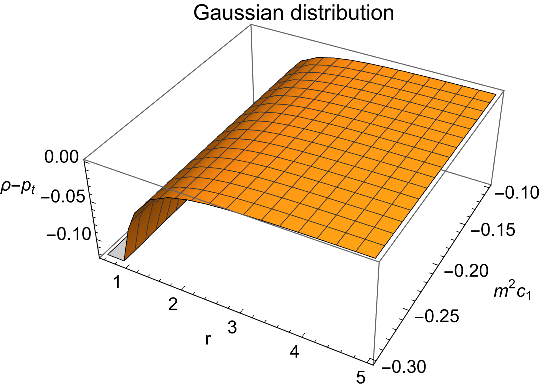}
        \caption{Energy conditions for Gaussian distribution. The values of the parameters are the same as Fig.~\ref{fig:1}. }
        \label{en1}
\end{figure}

\begin{figure}[htbp]
    \centering
    \includegraphics[scale=.6]{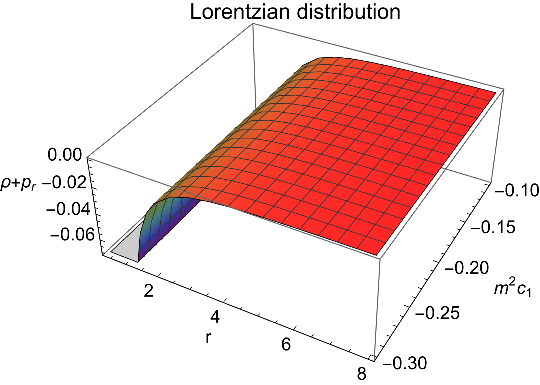}
    \includegraphics[scale=.6]{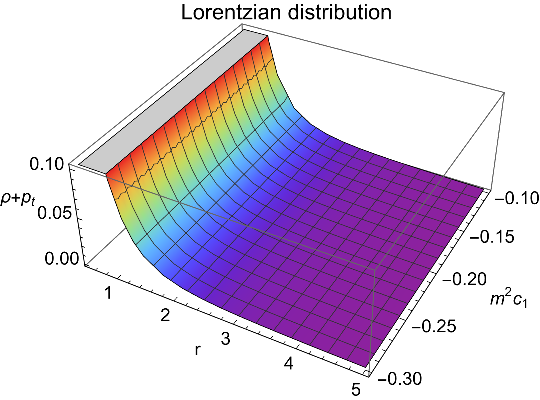}
    \includegraphics[scale=.6]{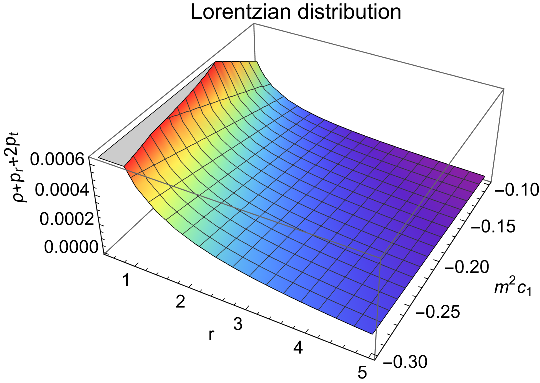}
    \includegraphics[scale=.6]{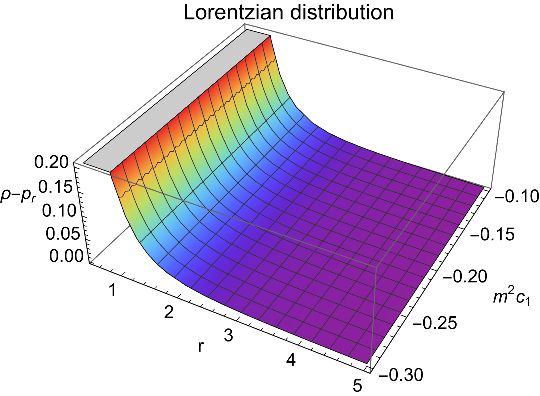}
    \includegraphics[scale=.6]{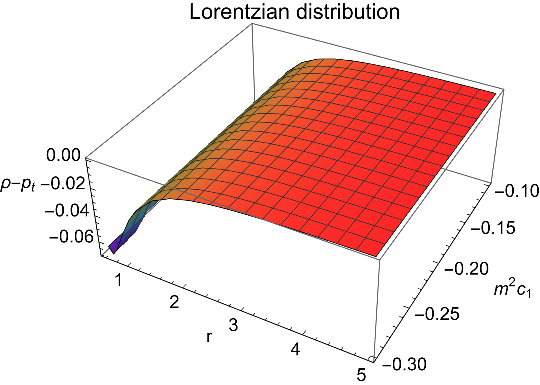}
        \caption{Energy conditions for Lorentzian distribution. The values of the parameters are the same as Fig.~\ref{fig:2}.  }
        \label{en2}
\end{figure}

The energy conditions become important when examining the properties of matter in the framework of modified gravity. The famous Raychaudhuri equations are the source of these conditions \cite{ad1,ad2,ad3}. Moreover, the purpose of all energy conditions is to investigate the geometry and composition of spherically symmetric spacetime. For an anisotropic matter distribution, the energy conditions are satisfied if the following inequalities hold:
 \begin{eqnarray*}
     (\mathcal{N E C}) &\Leftrightarrow& \rho+p_t \geq 0\; \text{and} \;\rho+p_r \geq 0.\\
(\mathcal{WEC}) &\Leftrightarrow& \rho \geq 0, \rho+p_t \geq \; \text{and} \;\rho+p_r \geq 0.\\
(\mathcal{SEC}) &\Leftrightarrow& \rho+p_j \geq 0\; \text{and} \;\rho+\sum_j p_j \geq 0, \forall j,\; \text{where} j=r, t.\\
(\mathcal{DEC}) &\Leftrightarrow& \rho \geq 0, \rho-\left|p_r\right| \geq 0\; \text{and} \;\rho-\left|p_t\right| \geq 0.
 \end{eqnarray*}
In our present study, with the help of graphical representation as shown in Fig. \ref{en1}-\ref{en2} we have checked that $\rho+p_r<0$, $\rho+p_t > 0$, $\rho+p_r+2p_t<0$, $\rho-p_t > 0$, and $\rho-p_t<0$ for both the wormhole models, and hence all the energy conditions are violated. The existence of the exotic matter in the throat of the wormhole is confirmed by the violation of the null energy criteria.

\subsection{Proper radial distance}
\begin{figure}[htbp]
    \centering
    \includegraphics[scale=.7]{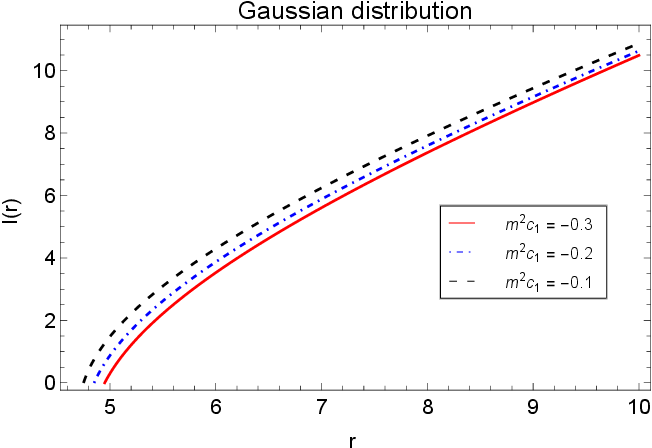}
    \includegraphics[scale=.7]{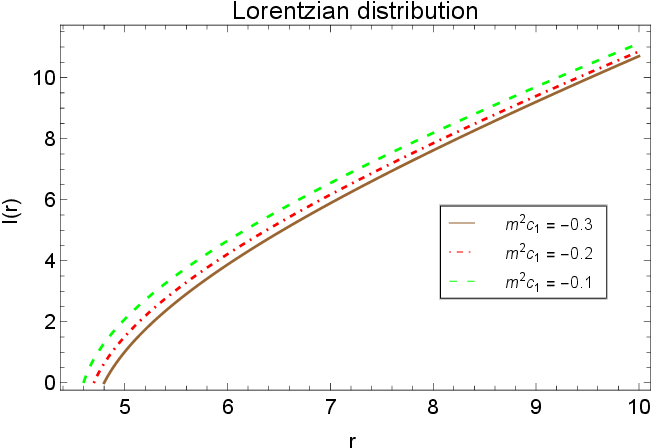}
        \caption{The proper radial distance of the wormhole. The values of the parameters are the same as Fig.~\ref{fig:1}-\ref{fig:2}.  }
        \label{lr}
\end{figure}

In this subsection, we are going to discuss the proper radial distance of $l(r)$ of the wormhole, which can be obtained as,
\begin{eqnarray}
    l(r)=\pm \int_{r_0+}^r\frac{1}{\sqrt{1-\frac{b(r)}{r}}}dr,
\end{eqnarray}
Due to the complex expression of $b(r)$, we performed the above integral numerically, and its profiles are shown in Fig~\ref{lr} for different values of $m^2c_1$. Fig.~\ref{lr} shows that $l(r)$ is finite everywhere and decreases from the upper universe $l(r)~\rightarrow~+\infty$ to the throat which is physically reasonable.

\subsection{Equilibrium condition by analyzing TOV}
\begin{figure}[htbp]
   \centering
   \includegraphics[scale=.75]{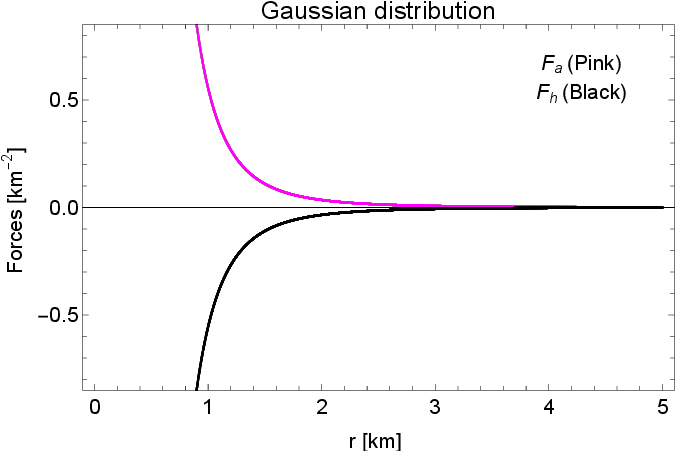}
    \includegraphics[scale=.75]{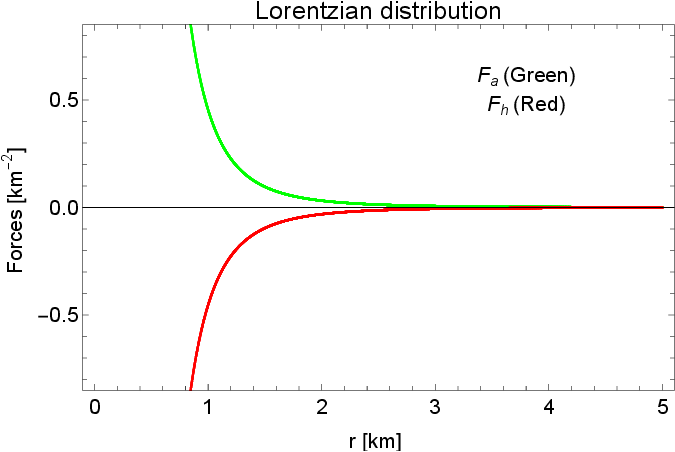}
        \caption{Two different forces acting on the wormhole spacetimes are shown in the figures. For drawing the profiles the values of the parameters used here are the same as Figs.~\ref{fig:1}-\ref{fig:2}. }
      \label{tov9}
\end{figure}

For our current work, the generalized Tolman-Oppenheimer-Volkov (TOV) equation, which is given by \cite{leon},
\begin{eqnarray}\label{tov1}
    \frac{dp_r}{dr}+\frac{2}{r}(p_t-p_r)+\frac{\phi'}{2}(\rho+p_r)=0,
\end{eqnarray}
provides the equilibrium condition.\\
The aforementioned equation is a significant and comprehensive technique that may be applied to investigate the stability conditions of astrophysical solutions, such as compact objects and wormholes.\\
Another way to write the Eq. (\ref{tov1}) is as
\begin{eqnarray}\label{tov2}
   F_g+F_a+F_h=0,
\end{eqnarray}
This describes the equilibrium condition of the wormhole.
Here,
\begin{eqnarray*}
    F_a&=&\frac{2}{r}(p_t-p_r),\\
    F_g&=&-\frac{\phi'}{2}(\rho+p_r),\\
    F_h&=&-\frac{dp_r}{dr}.
\end{eqnarray*}
where $F_g$ is the gravitational force, $F_h$ is the hydrostatic force, and $F_a$ is the force resulting from the anisotropic matter of the wormhole.
Nonetheless, it is evident from Eq. (\ref{tov2}) that the sum of three distinct forces must equal zero for the system to be in equilibrium. Fig.~\ref{tov9} displays the behavior of $F_a$, and $F_h$ for a specific selection of parameter values. However, since $\phi$ is assumed to be constant, the value of $F_g$ is $0$, meaning that the gravitational force does not influence our model. It is clear from the figure that the other two forces are precisely the same and opposing one another.
Therefore, it is clear that the three forces work together to establish the balance of the model, which supports the stability of the system.

\subsection{Measurement of exotic matter}
\begin{figure}[htbp]
   \centering
   \includegraphics[scale=.75]{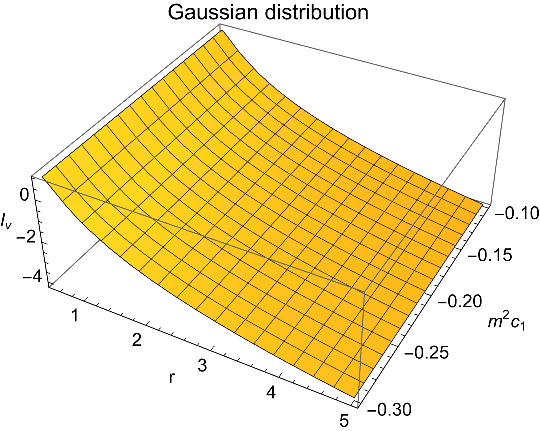}
    \includegraphics[scale=.75]{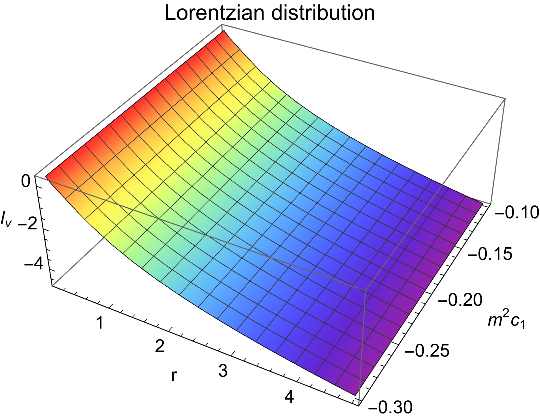}
        \caption{The VIQ are shown in the figures. For drawing the profiles, the values of the parameters used here are the same as Figs.~\ref{fig:1}-\ref{fig:2}. }
      \label{viq}
\end{figure}

Exotic matter is the term typically used to describe material that violates the NEC. The following volume integral quantifier (VIQ) can be computed to examine the amount of exotic matter near the wormhole's throat \cite{Visser:2003yf}:
\begin{eqnarray}
I_V=\oint(\rho+p_r)dV&=&-\int_{r_0}^{\infty}(1-b')\ln\left[e^{\nu}\left(1-\frac{b(r)}{r}\right)^{-1}\right]dr,
\end{eqnarray}
We can see how VIQ varies in connection with $r$ in Fig.~\ref{viq}. The model parameters are selected in this case as follows: For the Gaussian distribution the following values are taken: $C = 0.05,\, m^2c_2 = -0.05,\, M = 1.8,\,\theta = 0.02,$ and $E_1 = -1.05$, $m^2c_1~ \in[-0.3,\,0.1]$. For the Lorentzian distribution wormhole model, we select $C = 0.05,\, m^2c_2 = -0.05,\, M = 1.8,\, \phi = 0.02,\, E_2 = -1.05,$ and $m^2c_1 \in[-0.3,\,0.1]$.
In this connection, we want to mention that Lobo et al.  \cite{Lobo:2012qq} proposed an asymptotically flat phantom wormhole, and with the help of the volume integral quantifier, the author showed that as $\alpha$ tends to $-1$, the volume integral quantifier approaches to zero. Bhar \cite{Bhar:2023ruf} also used the volume integral quantifier to measure the amount of exotic matter to obtain the model of the wormhole in Rastall gravity. We can see from the figures that in both situations, there exists exotic matter near the throat of the WH since we have obtained $I_v<0$ for when r gets closer to $r_0$. This suggests that a very minimal amount of exotic matter is needed to open the throat of the wormhole.

\section{Photon deflection angle on null geodesics: Analyzing the deviation of photons' paths}\label{s4}
Let us delve into the investigation of the deviation angle of photons from null geodesics. To begin, we introduce a comprehensive line element that encompasses both spherically symmetric and static characteristics \cite{misner1973, schutz2014}. This line element can be expressed as follows:
\begin{equation}\label{eq38}
ds^2 = -\mathcal{A}(r) dt^2 + \mathcal{B}(r) dr^2 + \mathcal{C}(r) d\Omega^2.
\end{equation}
In order to analyze the trajectory of objects in free fall within this spacetime geometry \cite{schutz2014}, we utilize the geodesic equation. This equation establishes a connection between the momentum one-forms of the objects and the geometry of spacetime. Mathematically, it can be represented as:
\begin{equation}\label{eq39}
\frac{dp_\beta}{d\lambda} = \frac{1}{2} g_{\nu \alpha, \beta} p^\nu p^\alpha,
\end{equation}
 where $p^\alpha$ is the 4-momentum and $\lambda$ is the affine parameter. To examine the deviation angle of photons from null geodesics, we introduce the affine parameter $\lambda$ to characterize the motion. It is important to note that if the components of $g_{\alpha\nu}$ do not vary with respect to $x^\beta$ for a fixed index $\beta$, then $p_\beta$ remains constant throughout the motion. Focusing on the equatorial slice with $\theta = \pi/2$, we find that all components $g_{\alpha\beta}$ in Eq. \eqref{eq39} become independent of $t$, $\theta$, and $\phi$. As a result, we can identify the corresponding Killing vector fields $\delta^{\mu}\alpha \partial\nu$ with $\alpha$ as a cyclic coordinate. By denoting the constants of motion as $p_t$ and $p_\phi$, we can analyze the system's dynamics using the following expressions: $p_t = -E$,~and~$p_\phi = L$, where $E$ and $L$ represent the energy and angular momentum of the photon, respectively. With this in mind, we can express the geodesic equation as:
\begin{eqnarray}\label{eq41}
p_t = \dot{t} = g^{t \nu} p_\nu = E\mathcal{A}^{-1}(r), \\
p_\phi = \dot{\phi} = g^{\phi \nu} p_\nu = L\mathcal{C}^{-1}(r).
\end{eqnarray}
Here, the overdot notation indicates differentiation with respect to the affine parameter $\lambda$. Continuing from the previous analysis, we can derive the radial null geodesic as:
\begin{equation}\label{eq42}
\dot{r}^2 = E^2\mathcal{A}^{-1}(r)\mathcal{B}^{-1}(r) - L^2\mathcal{C}^{-1}(r)\mathcal{B}^{-1}(r).
\end{equation}

To provide a more precise formulation, let's express the equation for the photon trajectory in terms of the impact parameter $\eta = L/E$. The equation becomes:
\begin{equation}\label{eq43}
\left[ \frac{dr}{d\phi} \right]^2 = \eta^{-2}\mathcal{C}(r)^2 \mathcal{A}^{-1}(r)\mathcal{B}^{-1}(r) - \mathcal{C}(r) \mathcal{B}^{-1}(r)
\end{equation}

Now, we consider a photon source with a radius $r_s$ that affects the underlying geometry. To determine the deflection angle of the photons, we need to find the conditions under which the photons can reach the surface. This occurs when a solution $r_0$ satisfies the conditions $r_0 > r_s$ and $\dot{r}^2 = 0$, where $r_0$ represents the distance of the closest approach or the turning point. We can express the impact parameter as:
\begin{equation}\label{eq44}
\eta = \frac{L}{E} = \pm \sqrt{\mathcal{C}(r_0)\mathcal{A}^{-1}(r_0)}.
\end{equation}
In the weak gravity regime, we have $ \eta \approx \sqrt{\mathcal{C}(r_0)} $. Thus, if a photon originates from the polar coordinate limit, defined as $ \lim\limits_{r \to \infty} \left( r, -\frac{\pi}{2}-\frac{\alpha}{2} \right) $, passes through the turning point located at $ (r_0, 0) $, and approaches $ \lim\limits_{r \to \infty} \left( r, \frac{\pi}{2}+\frac{\alpha}{2} \right) $, we can define the deflection angle around wormhole throat as $ \alpha $. The deflection angle, denoted by $ \alpha $, depends on $r_0$ \cite{Bhattacharya:2010zzb}. It can be explicitly derived from Eq. (\ref{eq44}) as:
\begin{equation}\label{eq45}
\alpha(r_0) = -\pi + 2 \int_{r_0}^{\infty} \frac{\mathcal{B}^{1/2}(r) \mathcal{C}^{^{-1/2}}(r)}{\left[ \left( \mathcal{A}(r_0)\mathcal{A}^{-1}(r) \right) \left( \mathcal{C}(r)\mathcal{C}^{-1}(r_0) \right) -1 \right]^{1/2}} dr.
\end{equation}
\begin{figure}[htbp]
   \centering
   \includegraphics[scale=.65]{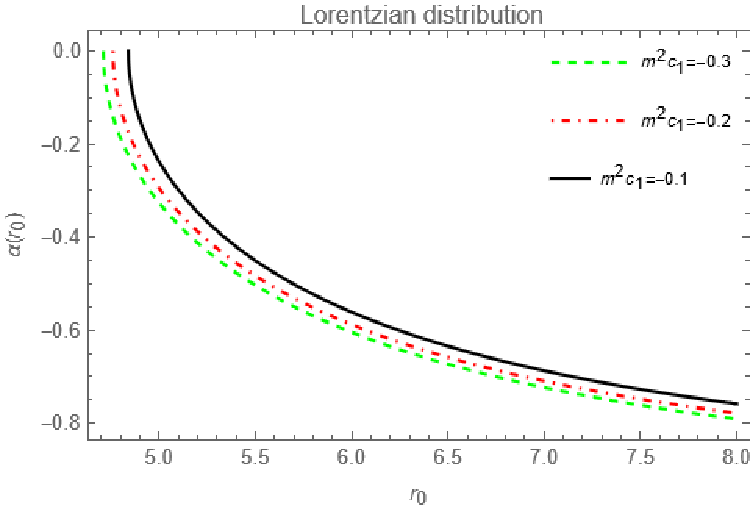}
     \includegraphics[scale=.65]{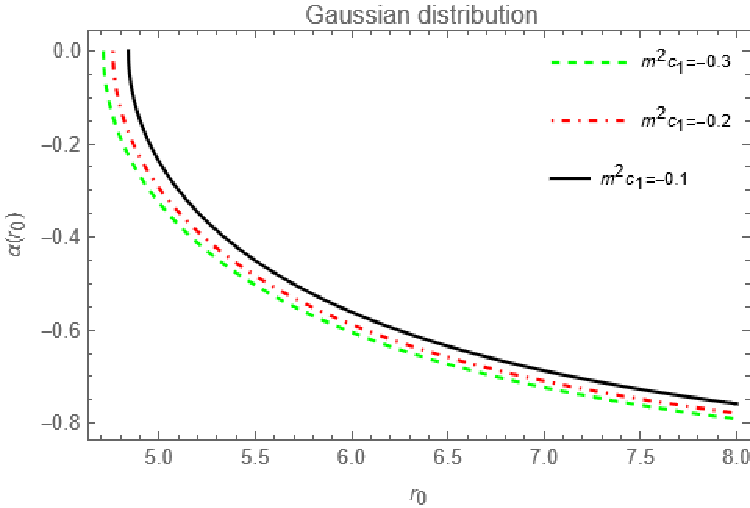}
      \caption{Deflection angle of photons in both scenarios. For drawing the profiles the values of the parameters used here are the same as Figs.~\ref{fig:1}-\ref{fig:2}. }
        \label{defGLD}
\end{figure}
By employing the specified metric coefficients within the wormhole geometry, it becomes straightforward to compute the deflection angle around the wormhole throat in dRGT massive gravity. This can be achieved through numerical integration of the aforementioned formulas (\ref{eq45}) while considering the shape functions detailed in Eqs. (\ref{sh1}) and (\ref{sh2}). The plots shown in Fig. \ref{defGLD} illustrate the phenomenon at hand. A negative deflection angle indicates the presence of repulsive gravity. The concept of the deflection angle around wormhole throat is introduced to validate the phenomenon of repulsive gravity. It is noteworthy that a negative deflection angle signifies the presence of repulsive gravity. Remarkably, the deflection angle consistently maintains negative values across all values of $r_0$ in both solutions. This consistent negativity serves as a manifestation of the repulsive gravity effect.

\section{Discussions} \label{s5}
In this work, we have provided the wormhole solution in the dRGT massive gravity, an extension of Einstein gravity. The effects of pressure anisotropy on the properties of the wormhole in the presence of space non-commutativity have been investigated. The unique feature of this noncommutative geometry is that it substitutes the point-like structure of a gravitational source by the smeared distribution of the energy density. To prevent the appearance of event horizons, we simplify our calculations by assuming a constant redshift function, i.e., $\phi' = 0$. We solved the field equations in this modified gravity, taking into account two distinct matter densities, and derived the pressure profiles and shape functions. By carefully selecting the constant terms, we have also drawn the wormhole shape function, which is monotonically increasing. Interestingly, as $m^2c_1$ increases, the throat radius for both models decreases. Flare-out conditions have also been met for the various choices of $m^2c_1$ shown in the figure, which is another crucial criterion for wormhole geometry. Since the asymptotic flatness is lost, i.e., $b(r)/r$ does not tend to zero as $r$ tends to $\infty$, the shape function violates the Morris-Thorne characteristics. With the assistance of the graphical illustration, the NEC and all other energy criteria have been checked, and it is observed that for both models, the NEC is violated and it is necessary to keep the wormhole throat open. This is because of the exotic anisotropic dark energy nature of coupled curvature and massive gravitons. The shape function provides the embedding function of the wormhole and, when rotated along the z-axis, provides the complete visualization of the wormhole structure, which has been studied in detail. The Tolman-Oppenheimer-Volkov (TOV) equation is also examined for the study of the equilibrium conditions in both cases. It is seen that anisotropic and hydrostatic forces interact to keep the wormhole model stable even though the gravitational force is zero because of the constant redshift function. Our investigation extended beyond the TOV equation to explore the repulsive nature of gravity. We discovered that the presence of repulsive gravity is associated with a negative deflection angle. To validate this phenomenon, we introduced the concept of the photon deflection angle on the wormhole. Notably, we observed that the deflection angle consistently maintains negative values across all values of $r_0$ in the two solutions. This consistent negativity can be interpreted as a manifestation of the repulsive gravity effect. In this respect, we would like to draw attention to some previous research on wormhole geometries where non-commutative geometry is involved. In \cite{Rahaman:2014dpa}, a novel wormhole solution influenced by noncommutative geometry was examined. It included the requirement to permit conformal Killing vectors and to assume a Lorentzian distribution of particle-like gravitational source. In $f(T,\,T_G)$ gravity, the noncommutative wormhole solution was obtained in \cite{Mustafa:2019eet}. Chalavadi et al. \cite{Chalavadi:2023zcw} obtained the wormhole solution in $f(Q,T)$ gravity by choosing a linear form of $f(Q,T)=\alpha Q+\beta T$, whereas Debnath et al. \cite{Debnath:2023yfm} choose $f(R)$ gravity, Rahaman et al. \cite{Rahman:2022pug}
 used Einstein-Gauss-Bonnet gravity, and Feng \cite{Feng:2022zlc} used exponential gravity to obtain the model of the wormhole in the context of noncommutative geometry. All of them described the effect of the coupling parameter of the modified gravity in the spacetime of the wormhole but we are the first to describe the effect of the graviton mass on the wormhole spacetime in the presence of noncommutative geometry. Finally, our analysis shows that there exist traversable wormhole solutions having physically intriguing features with zero tidal force that have both a Gaussian and Lorentzian distribution for the matter density in the realm of dRGT massive gravity.

\section*{acknowledgments}
PB is thankful to IUCAA, Government of India for providing visiting associateship. AE thanks the National Research Foundation of South Africa for the award of a postdoctoral fellowship.

\end{document}